\documentclass[aps,nofootinbib,prd,preprintnumbers]{revtex4}
\pdfoutput=1
\usepackage{amstext}
\usepackage{amssymb}
\usepackage{amsmath}
\usepackage{graphicx}
\usepackage{rotating}
\usepackage{gensymb}
\usepackage[format=plain,singlelinecheck = false, format= hang, justification=raggedright]{caption}

\usepackage{multirow} % To use multirow command in tables
\usepackage[T1]{fontenc} 
\usepackage{slashed}
\usepackage{caption}
\usepackage{float}
\usepackage[normalem]{ulem}

\usepackage[usenames,dvipsnames]{color}
\definecolor{mightnightblue}{RGB}{25,25,112}
\usepackage{hyperref}
 \hypersetup{
     colorlinks=true,
     linkcolor= Black,
     citecolor=mightnightblue,
     urlcolor=mightnightblue
     } 
\def\gsim{\raise0.3ex\hbox{$\;>$\kern-0.75em\raise-1.1ex\hbox{$\sim\;$}}}
\def\lsim{\raise0.3ex\hbox{$\;<$\kern-0.75em\raise-1.1ex\hbox{$\sim\;$}}}
\newcommand {\ignore}[1]{}

\newcommand{\ie} {{\it i.e.}}
\newcommand{\eVq}  {\text{eV}^2}
\newcommand{\be}{\begin{equation}}
\newcommand{\ee}{\end{equation}}
\newcommand{\bea}{\begin{eqnarray}}
\newcommand{\eea}{\end{eqnarray}}

\newcommand{\pme}{P_{\mu e}}

\newcommand{\pmm}{P_{\mu \mu}}

\newcommand{\dcp}{\ensuremath{\delta_{\text{CP}}}}

\newcommand{\sstc}{\ensuremath{\sin^{2}\theta_{23}}}

\newcommand{\aet}{\ensuremath{a_{e\tau}}}
\newcommand{\amt}{\ensuremath{a_{\mu\tau}}}
\newcommand{\att}{\ensuremath{a_{\tau\tau}}}
\newcommand{\aee}{\ensuremath{a_{ee}}}
\newcommand{\aem}{\ensuremath{a_{e\mu}}}
\newcommand{\amm}{\ensuremath{a_{\mu\mu}}}

\newcommand{\AddrAHEP}{%
  AHEP Group, Institut de F\'{i}sica Corpuscular --
  CSIC/Universitat de Val\`{e}ncia, Parc Cientific de Paterna.\\
 C/ Catedratico Jos\'e Beltr\'an, 2 E-46980 Paterna (Val\`{e}ncia) - SPAIN}

\begin{document}

\begin{flushright}
FTUV-18-05-31 \\
IFIC-18/23 
\end{flushright}

\title{Exploring the intrinsic Lorentz--violating parameters at DUNE}

\author{Gabriela Barenboim$^1$}\email{Gabriela.Barenboim@uv.es}
\author{Mehedi Masud$^2$}\email{masud@ific.uv.es}
\author{Christoph A. Ternes$^2$}\email{chternes@ific.uv.es}
\author{Mariam T{\'o}rtola~$^2$}\email{mariam@ific.uv.es}  
\affiliation{$^1$~Departament de F{\'i}sica Te{\'o}rica and IFIC, Universitat de Val{\`e}ncia-CSIC, E-46100, Burjassot, Spain}
\affiliation{$^2$~\AddrAHEP}
 
\begin{abstract}

Neutrinos can push our search for new physics to a whole new level. What makes them so hard to be detected, what allows them to travel humongous distances without being stopped or deflected allows to amplify Planck suppressed effects (or effects of comparable size) to a level that we can measure or bound in DUNE.
In this work  we analyze the sensitivity of DUNE to CPT and Lorentz--violating interactions in a framework that allows a straightforward extrapolation of the bounds obtained to any phenomenological modification of the dispersion relation of neutrinos. 
\end{abstract}

%\pacs{14.60.Pq,13.15.+g,12.60.-i} 
\maketitle

%\newpage

%~~~~~~~~~~~~~~~~~~~~~~~~~~~~~~~~~~~~~~~~~~~~~~~~~~~~~~~~~~~~~~~~~~~~~~~~~~~~~~~~~~~~~~~~~~~~~~~~~~~~~~~%

\section{Introduction}
\label{sec:intro}

The Standard Model (SM) of particle physics (including neutrinos) is arguably the most
successful and better tested theory in fundamental physics. Most of our experimental
efforts today and the planned ones for 
the future are driven by the search of whatever physics hides
behind it. In lack of a solid theoretical guidance, all avenues should be explored
and all the possible paths followed. No corner should be left unattended.

Our way of describing Nature, the tool that allowed us to get to this point, local 
relativistic quantum field theory should not avoid the scrutiny. Ways need to be found
to test also our dearest tools and the most sacred cows derived from or associated to them. 
Massive neutrinos are an ideal engine for such a purpose.
The origin of neutrino masses lies probably beyond the SM and is associated with a new scale
where new physics is expected to play a role and therefore oscillation experiments offer a window where
our journey can start.

The CPT theorem~\cite{Streater:1989vi} is undoubtedly the prodigal son of local relativistic quantum field theories.
Such a status is derived from the fact that the theorem is based only on three sensible assumptions, Lorentz invariance, Hermiticity of the Hamiltonian and locality, all which are incorporated for reasons beyond holding CPT itself.  
CPT violation can be incorporated in
field theory by giving up locality but not the other two~\cite{Barenboim:2002tz}, and will
be tested to an unprecedented level by future experiments~\cite{Barenboim:2017ewj}\footnote{{Ref.~\cite{Barenboim:2017ewj} focus  on a model--independent way to study CPT violation in the neutrino sector, see also Ref.~\cite{deGouvea:2017yvn}. However, there might be different forms of CPT violation which do not affect neutrinos at all. }}. If 
CPT violation exists and is related to quantum gravity, which is clearly non--local and expected to
be suppressed by the Planck mass, long--baseline experiments will have it within their reach.

Even though such experiments are significantly sensitive to
new physics in neutrino propagation, unfortunately, they involve long journeys through the Earth crust and therefore 
incorporate important matter effects which induce a fake (non--fundamental) 
CPT violation.
An attempt at disentangling matter effects from genuine CPT asymmetries in neutrino oscillation for various baselines and energies can be found in \cite{Rout:2017udo}. Discrimination of true CPT violation, a violation that challenges our description
of Nature in terms of local relativistic quantum field theories, from non--standard interactions 
which can induce a difference between matter and antimatter but do not imply a change of paradigm 
is also crucial and was analyzed in \cite{Barenboim:2018lpo}.

Other forms of CPT violation are to be tested as well. Especially interesting is a potential
spontaneous CPT violation associated to a Lorentz invariance breaking.
As we have stated before, the CPT theorem holds only in flat space time if and only if the 
theory respects locality, Hermiticity and also Lorentz invariance. A breakdown of Lorentz invariance is normally linked with quantum gravity and  the existence of a universal fundamental
length scale. Such a universal (i.e. common to all frames) scale would collide with general relativity, as Lorentz transformations predict length contractions. A way out of this apparent contradiction
is a modification of Lorentz transformations which can be incorporated in the model as modified dispersion relations (or nonlinear Dirac equations). This fact is especially interesting for several reasons, as will be seen below.

The  usually identified as the most stringent bound on CPT violation comes  from the neutral kaon system~\cite{PDG}, 
\begin{equation}
  %|m(K^0) - m(\overline{K}^0)| < \; 0.6 \times 10^{-18} m_K \; \approx \; 0.3 \times 10^{-9}\, \mbox{eV} \,. 
  \frac{|m(K^0) - m(\overline{K}^0)|}{m_K} < \; 0.6 \times 10^{-18}  \,.
  \label{eq:mK}
\end{equation}
However,  one can notice that the strength of this bound is relative. First of all, 
using the kaon mass as a scale is somewhat arbitrary, because we do not know at which scale effects of CPT violation should arise.
Besides that, since kaons are bosons and in this case  the masses squared are the quantities  that enter the Lagrangian, it could be more appropriate expressing the limit above in terms of the squared masses.  
Following this choice, it has been found that neutrinos can provide a very good bound on CPT violation~\cite{Barenboim:2017ewj}, 
\begin{eqnarray}
 & |\Delta m_{21}^2-\Delta \overline{m}_{21}^2| &< 4.7\times 10^{-5} \, \text{eV}^2,
  \nonumber \\
  & |\Delta m_{31}^2-\Delta \overline{m}_{31}^2| &< 3.7\times 10^{-4} \, \text{eV}^2.
 \end{eqnarray} 
These limits have been derived using neutrino oscillation data, insensitive to the masses themselves, but providing very precise measurements of the neutrino mass squared differences.
Note as well that the choice of this observable is also motivated by the fact that the mass squared is the parameter entering the dispersion relation $E^2 = p^2 + m^2$  
and the natural parameter in relativistic kinematics. Therefore, deviations from this standard dispersion relation are a way to explore CPT violations triggered by Lorentz invariance breaking.

Clearly, there are multiple ways to modify the dispersion relation and most of them can be easily incorporated into the Hamiltonian and bounded. This is precisely what we are going to do in this work.  We therefore
classify the possible modifications of the dispersion relation in a general (agnostic) way according 
to the spectral distortions they give to the oscillation pattern, (\ie \, their dependence in energy) and
bound those that can mimic non--standard neutrino interactions, namely those which behave as matter effects.

We will first  illustrate with an example how the transition probability gets affected by a 
modification of a dispersion relation. 
Let us assume the neutrino energy is given by
\begin{equation}
E^2 = p_i^2 + \frac{1}{2} m_i^2   \left( 1+ e^{2 A_i E /m_i^2}\right),
\end{equation}
where $m_i $ is the neutrino mass, $E$ is its energy, $p_i$ its momentum and $A_i$ is a dimensionful
and non--universal Lorentz breaking parameter. 
We will assume  neutrinos are produced with the same energy and we will
restrict ourselves to the case of two generations for simplicity. 
In this case, the transition probability is given by,
\begin{equation}
 P_{\alpha\beta}=1-\sin^22\theta\sin^2\left(\frac{\Delta p \; L}{2}\right),
\end{equation}
where $\theta $ is the mixing angle. The neutrino momentum $p_i$ can be written as 
\begin{equation}
p_i = \sqrt{E^2 - \frac{1}{2} m_i^2   \left( 1+ e^{2 A_i E /m_i^2}\right)} 
 \approx   E - \frac{1}{4 E }   \left( \ m_i^2 + m_i^2 e^{2 A_i E /m_i^2} \right),
\end{equation}
and therefore
\begin{equation}
\Delta p  \approx \frac{1}{4 E }   \left( \Delta m^2 + m_\alpha^2 e^{2 A_\alpha E /m_\alpha^2}  
 - m_\beta^2 e^{2 A_\beta E /m_\beta^2}  \right) 
 \approx   \frac{\Delta m^2 }{2 E }  + \frac{1}{2}(A_\alpha - A_\beta).
\end{equation}
Clearly, neutrino oscillation experiments offer a unique opportunity to test this kind of physics.
It should be noticed, however, that only non--universal \ie \, family dependent modifications are accessible to
oscillations. Universal parameters ($A=A_\alpha=A_\beta$) give rise to an oscillation pattern which possesses not only a different spectral shape but also is quadratic on the already small parameter $A$
\begin{eqnarray}
\Delta p \approx \frac{\Delta m^2 }{2 E } - \frac{1}{2} \frac{ A^2 \; \Delta m^2 \; E}{m_\alpha^2 \, m_\beta^2}\, .
\end{eqnarray}

Existing limits on the neutrino mass squared differences from atmospheric and long--baseline neutrino experiments have already pushed the limits down 
to the ballpark where we can expect effects of string theory or quantum gravity to show up.
Thus, DUNE, with its improved sensitivity to the atmospheric $\Delta m^2$, will be able to explore the regions
where these new scales, associated with physics at a scale we have been so far only able
to speculate with, leave trace.

%%%%%%%%%%%%%%%%%%%%
\section{Theoretical Background}
\label{sec:theo}
%%%%%%%%%%%%%%%%%%%%

Lorentz--violating neutrinos and antineutrinos are effectively described by the 
Lagrangian density~\cite{Kostelecky:2003cr, Kostelecky:2011gq}
\be\label{eq:1}
\mathcal{L} = \frac{1}{2}\bar{\Psi}({i\slashed{\partial}} - M + \hat{\mathcal{Q}})\Psi + h.c.
\ee
where $\hat{\mathcal{Q}}$ is a generic Lorentz--violating operator.
There exists a subset of Lorentz--violating operators that also break CPT invariance in the fermion sector.
Restricting our attention only to the renormalizable Dirac couplings of the theory, we can  start from the Lorentz--violating Lagrangian~\cite{Kostelecky:2011gq}
\be\label{eq:lag_cpt}
\mathcal{L}_{\text{LIV}} = -\frac{1}{2}\left[ a^{\mu}_{\alpha\beta}\bar{\psi}_{\alpha}\gamma_{\mu}\psi_{\beta} + b^{\mu}_{\alpha\beta}\bar{\psi}_{\alpha}\gamma_{5}\gamma_{\mu}\psi_{\beta} 
- i  c_{\alpha\beta}^{\mu\nu}   \bar{\psi}_{\alpha}\gamma_{\mu}\partial_\nu\psi_{\beta}
- i d_{\alpha\beta}^{\mu\nu}   \bar{\psi}_{\alpha}\gamma_5\gamma_{\mu}\partial_\nu\psi_{\beta}
\right] + h.c.
\ee
The observable effect on the left handed neutrinos is controlled by the combinations
\be
(a_{L})^{\mu}_{\alpha\beta} = (a + b)^{\mu}_{\alpha\beta} \, , \quad \quad  (c_{L})^{\mu\nu}_{\alpha\beta} = (c + d)^{\mu\nu}_{\alpha\beta}  \, ,
\ee
 which are constant hermitian matrices in the flavor space that can modify the standard vacuum Hamiltonian. The first combination is relevant for CPT--violating neutrinos, whereas the second combination is only relevant for CPT-even Lorentz-violating neutrinos.
 In this work we will focus on the isotropic component
 of the Lorentz--violating terms, and therefore we will fix the ($\mu$,$\nu$)  indices to 0.
To simplify our notation, from now on, we will denote the parameters $(a_{L})^{0}_{\alpha\beta}$ and   $(c_{L})^{00}_{\alpha\beta}$ as $a_{\alpha\beta}$ and $c_{\alpha\beta}$\footnote{{
These components are  defined in the Sun--centered celestial equatorial frame~\cite{Kostelecky:2003cr}.
}}.

Explicitly, one can write the Lorentz--violating contribution to the full oscillation Hamiltonian
\be\label{eq:h_cpt}
H = H_{\text{vac}} + H_{\text{mat}} + H_{\text{LIV}}\, ,
\ee
 as~\cite{Diaz:2015dxa}
 \be\label{eq:cpt_part}
 H_{\text{LIV}} = 
 \left(
 \begin{array}{ccc}
a_{ee} & a_{e\mu} & a_{e\tau} \\
a^*_{e\mu} & a_{\mu\mu} & a_{\mu\tau} \\
a^*_{e\tau} & a^*_{\mu\tau} & a_{\tau\tau}
\end{array}
\right)
-\frac{4}{3}
 E \left(
 \begin{array}{ccc}
c_{ee} & c_{e\mu} & c_{e\tau} \\
c^*_{e\mu} & c_{\mu\mu} & c_{\mu\tau} \\
c^*_{e\tau} & c^*_{\mu\tau} & c_{\tau\tau}
\end{array}
\right)
\, .
\ee
Note that the the effect  of $a_{\alpha\beta}$--induced Lorentz violation in neutrino oscillations is proportional to the neutrino baseline $L$, while the terms corresponding to $c_{\alpha\beta}$ induce new contributions proportional to $L E$. In the latter, the factor $-4/3$ arises from the non--observability of the Minkowski trace of  $c_L$, which forces the components $xx$, $yy$, and $zz$ to be related to the 00 component~\cite{Kostelecky:2003cr}.
Anyway, in this work we will consider only the presence of Lorentz--violating terms generated by the first type of terms.
Updated constraints on $c_{\alpha\beta}$, mainly  from Super--Kamiokande, can be found in Refs.~\cite{Abe:2014wla,Kostelecky:2008ts}.

Then, if we focus  on the first term in Eq.~(\ref{eq:cpt_part}), we notice that this Hamiltonian looks very similar to the one corresponding to nonstandard interactions (NSI) in the
neutrino propagation
\be\label{eq:h_nsi}
H^{'} = H_{\text{vac}} + H_{\text{mat}} + H_{\text{NSI}}\, ,
\ee
where the NSI term is parametrized as
\be\label{eq:nsi_part}
H_{\text{NSI}} = 
\sqrt{2}G_{F} N_{e}\left(
 \begin{array}{ccc}
\epsilon^m_{ee} & \epsilon^{m}_{e\mu} & \epsilon^{m}_{e\tau} \\
\epsilon^{m}_{\mu e} & \epsilon^{m}_{\mu\mu} & \epsilon^{m}_{\mu\tau} \\
\epsilon^{m}_{\tau e} & \epsilon^{m}_{\tau\mu} & \epsilon^{m}_{\tau\tau}
\end{array}
\right) 
\, .
\ee
Here $N_e$ corresponds to the electron number density along the neutrino trajectory and the parameters $\epsilon^m_{\alpha \beta}$ give the relative strength between NSI matter effect in a medium and the Standard Model weak interactions. 
One thus finds a correlation between the NSI and CPT--violating scenario through the following relation~\cite{Diaz:2015dxa},
\be\label{eq:nsi_cpt}
a_{\alpha\beta} \equiv \sqrt{2}G_{F} N_{e}\epsilon^{m}_{\alpha\beta}\, .
\ee
However, there are important differences between these two  scenarios. NSI during neutrino propagation is basically an exotic matter effect and, hence, plays no role in vacuum, whereas the type of  CPT  violation considered here is an intrinsic effect, present even in vacuum. Nevertheless, the  equivalence in Eq.~(\ref{eq:nsi_cpt}) allows the study of the CPT--violating parameters in long baseline experiments in an approach similar to the treatment of NSI in neutrino propagation.
In the following, we estimate the sensitivity on the CPT--violating parameters with the help of DUNE~\cite{Acciarri:2015uup}, showing the correlations among the Lorentz--violating and oscillation parameters as well.
We will also compare our results with the limits in the literature.

\section{Numerical procedure and simulation of DUNE} 
\label{sec:num}

The Deep Underground Neutrino Experiment (DUNE) will consist of two detectors exposed to a megawatt-scale muon neutrino beam
produced at Fermilab. The near detector will be placed close to the source of the beam, while the second detector, 
comprising four 10 kton liquid argon TPCs will be installed 1300 kilometers away from the neutrino source in the Sanford Underground 
Research Facility in South Dakota. The primary scientific goals of 
DUNE include the measurement of the amount of leptonic CP violation, the determination of the neutrino mass ordering and the precision 
measurement of the neutrino mixing parameters.
To simulate DUNE we use the GLoBES package \cite{Huber:2004ka,Huber:2007ji} with the most recent DUNE configuration file provided by 
the Collaboration \cite{Alion:2016uaj} and used to produce the results in Ref.~\cite{Acciarri:2015uup}. We assume DUNE to be running 3.5 years 
in the neutrino mode and another 3.5 years in  the antineutrino mode. Assuming an 80 GeV beam with 1.07 MW beam power, this corresponds to 
an exposure of 300 kton--MW--years. 
In this configuration, DUNE will be using $1.47\times 10^{21}$ protons on target (POT) per year, which  basically amounts in one single
year to the same statistics  accumulated by T2K in runs 1-7c~\cite{Abe:2017uxa}. Our analysis includes disappearance and appearance 
channels, simulating signals and backgrounds. The simulated backgrounds include contamination of antineutrinos (neutrinos) in the neutrino
(antineutrino) mode, and also misinterpretation of flavors, as discussed in detail in \cite{Alion:2016uaj}.

To analyze the Lorentz--violating scenario, we perform our simulation of the DUNE experiment using the GLoBES--extension 
\textit{snu.c}, presented in Refs.~\cite{Kopp:2006wp, Kopp:2007ne}. This extension was originally made to include nonstandard 
neutrino interactions and sterile neutrinos in GLoBES simulations. For this analysis we have modified  the definition of the neutrino oscillation probability function inside 
 \textit{snu.c}, implementing the Lorentz--violating Hamiltonian in Eq.~(\ref{eq:cpt_part}).

The DUNE fake data sample is simulated using the standard oscillation parameters in Tab.~\ref{tab:oscpar}, taken from Refs.~\cite{deSalas:2017kay,globalfit}.  For the neutrino mass ordering, we 
 assume a normal ordered spectrum. Then, we try to reproduce the simulated data with the CPT--violating Hamiltonian in Eq.~(\ref{eq:h_cpt}), allowing for Lorentz violation. 
For simplicity, in our statistical analysis, we consider only one  or two Lorentz--violating parameters different from zero at the time. Since DUNE has no sensitivity  to the solar parameters and since  
 $\theta_{13}$ is rather well measured by current reactor and long--baseline experiments, we keep these values fixed to their current best fit value, while marginalizing 
over  $\theta_{23}$ and $\delta$ , if not plotting them. In addition, we marginalize over the atmospheric mass splitting, $\Delta m_{31}^2$, allowing for the two possible mass orderings. 
When studying a non--diagonal Lorentz--violating parameter, $a_{\alpha\beta}$, we also marginalize 
over its corresponding phase, $\phi_{\alpha\beta}$. Therefore, if we study two non--diagonal complex parameters simultaneously, we marginalize over a total of five parameters. 

\begin{table}[t!]\centering
  \catcode`?=\active \def?{\hphantom{0}}
   \begin{tabular}{lc}
    \hline  \\[-3mm] 
    parameter & value 
    \\ \\[-3mm] 
    \hline\\[-3mm]  
    $\Delta m^2_{21}$& $7.55\times 10^{-5} \, \eVq$\\  \\[-3mm]  
    $\Delta m^2_{31}$&  $2.50\times 10^{-3} \, \eVq$\\ \\[-3mm]  
    $\sin^2\theta_{12}$ & 0.320\\  \\[-3mm]  
     $\sin^2\theta_{23}$ &  0.547\\ \\[-3mm]  
    $\sin^2\theta_{13}$ & 0.02160\\ \\[-3mm]  
   $\delta$ & -0.68$\pi$\\
       \hline
     \end{tabular}
       \captionsetup{justification=centering}
     \caption{The standard oscillation parameters used throughout this work~\cite{deSalas:2017kay,globalfit}.}
     \label{tab:oscpar} 
\end{table}

%%%%%%%%%%%%%%%%%%%
\section{Results}
\label{sec:results}
%%%%%%%%%%%%%%%%%%%

\begin{figure}[t!]
\centering
\includegraphics[scale=0.54]{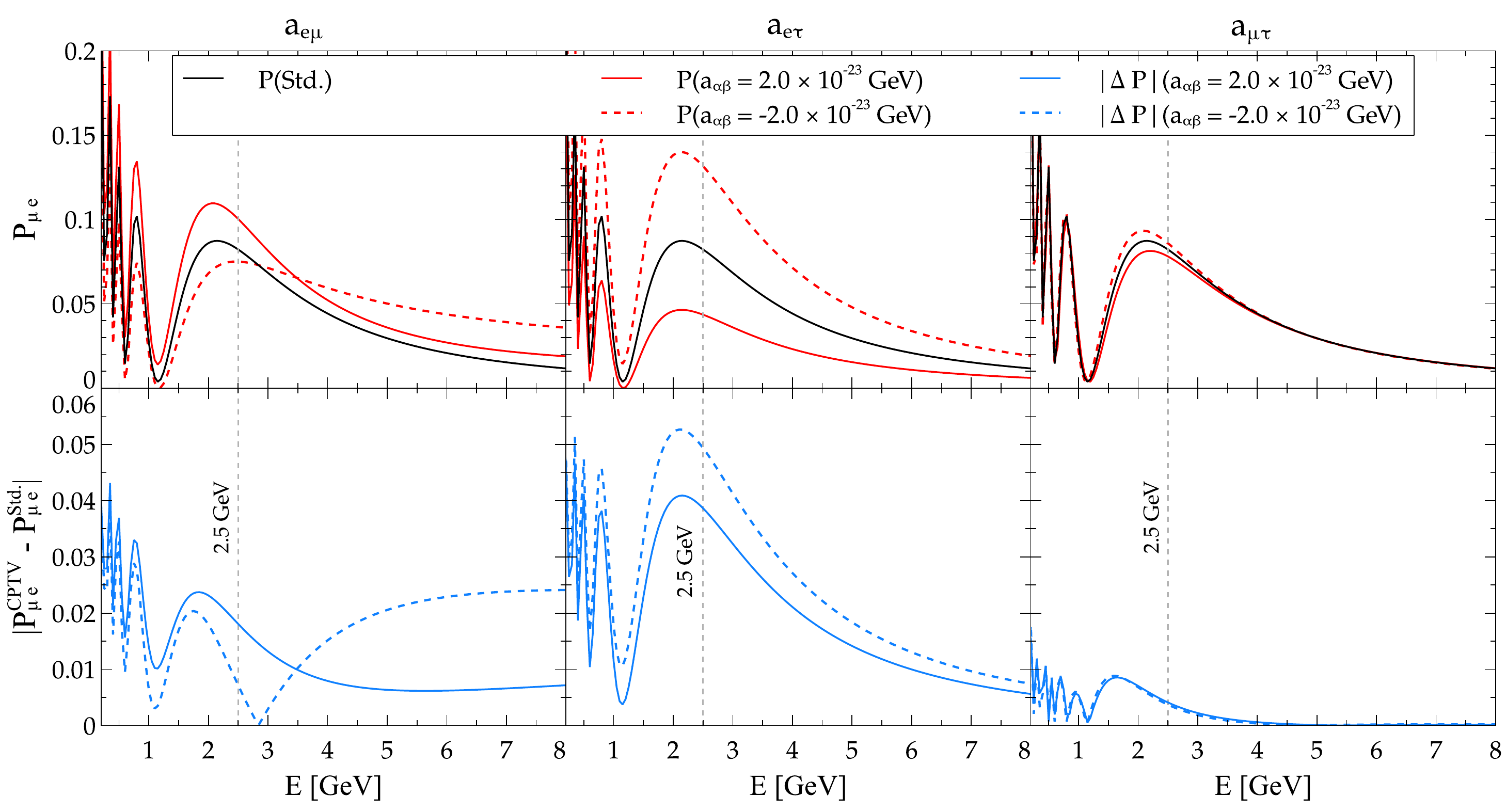}
\caption{\footnotesize{
Top: oscillation probability for the $\nu_{\mu} \to \nu_{e}$ channel  as a function of energy with standard matter effect (black) and with a non--diagonal CPT--violating parameter (red). 
Bottom: absolute difference between the standard and CPT--violating oscillation probabilities shown above.
Each column corresponds to a different
non--diagonal CPT--violating parameter taken  different from zero and fixed to $\pm 2.0 \times 10^{-23}$ GeV. The DUNE baseline (1300 km) has been assumed, while the peak energy of the neutrino flux in DUNE is indicated by a vertical line.}}
\label{fig:prob_e_aij}
\end{figure}

\begin{figure}[h!]
\centering
\includegraphics[scale=0.54]{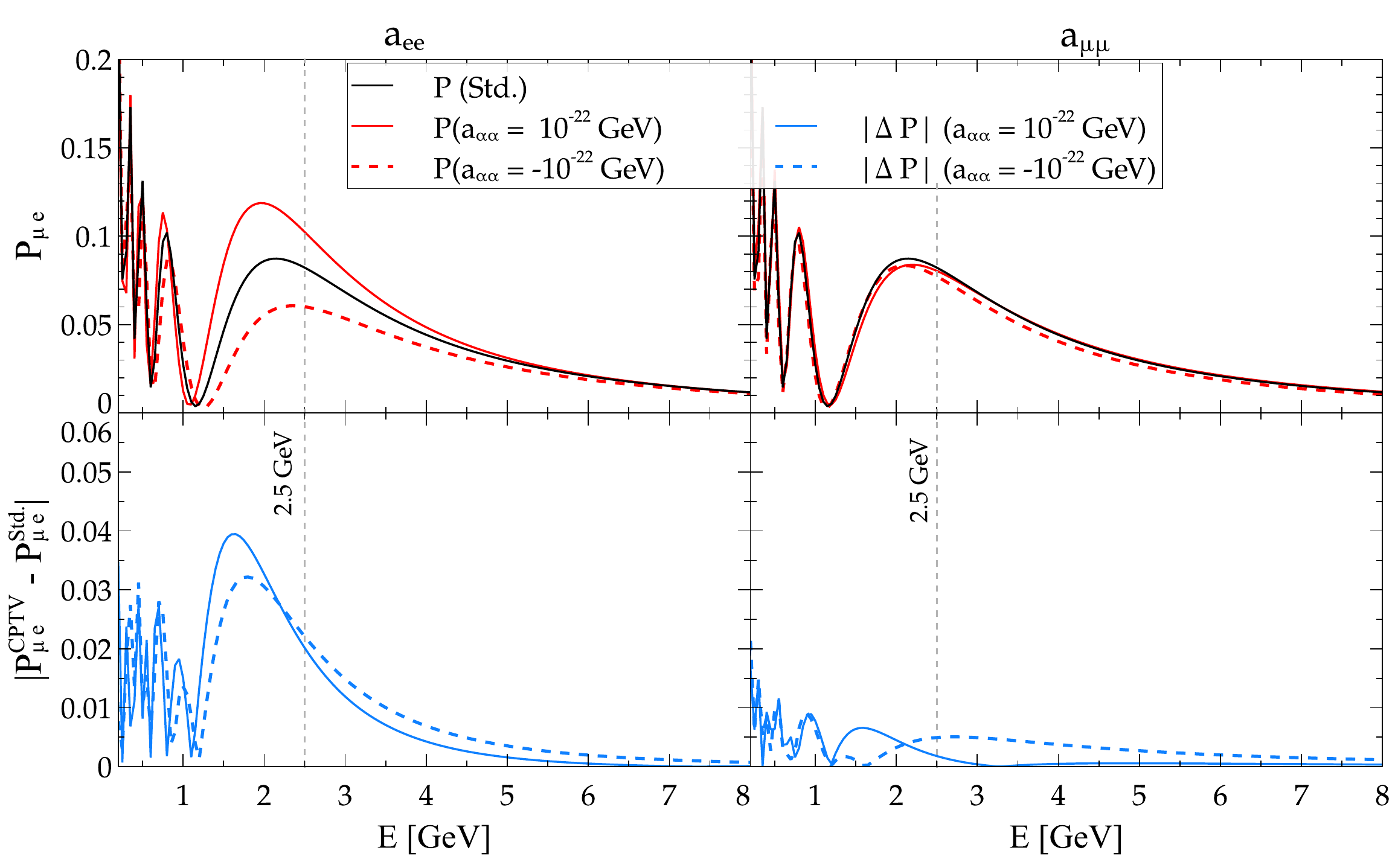}
\caption{\footnotesize{Same results as in Fig.~\ref{fig:prob_e_aij}, now for the diagonal parameters $\aee$ and $\amm$. 
Note, however, that the benchmark values used for the diagonal CPT--violating parameters here are five times larger than the ones assumed in Fig.~\ \ref{fig:prob_e_aij}.}}
\label{fig:prob_e_aii}
\end{figure}

In this section we present the results obtained in our analysis. 
First of all, we study the impact of CPT violation at the probability level, see Figs.~\ref{fig:prob_e_aij} and \ref{fig:prob_e_aii}.
In order to understand the role of the individual CPT--violating parameters, $a_{\alpha\beta}$, here we consider one of them different from zero at a time. 
To illustrate the effect at the probability level, we have  focused on the dominant channel at the  peak of the
DUNE neutrino flux, namely the appearance channel $\nu_{\mu} \to \nu_{e}$. The DUNE neutrino flux  is expected to peak around 
$2-4$ GeV. Note that $\pme$  has a peak around that energy while $\pmm$ has a dip. 
This enables DUNE to get a dominant contribution from the $\nu_{\mu} \to \nu_{e}$ channel. 
Nevertheless, we have considered the appearance and disappearance channels for our statistical analyses.\\

From Fig.~\ref{fig:prob_e_aij}, we note that $\aem$ and $\aet$ have a larger impact on the oscillation probability  $\pme$ compared to $\amt$. 
One can also see that the presence of a non--zero $\aem$ parameter may give rise to spectral distortions in the neutrino appearance probability.
On the other hand, a positive (negative) value of the parameter $\aet$ would produce a shift of the probability to smaller (larger) values. We have considered positive and negative values for these parameters in Fig.~\ref{fig:prob_e_aij}, although the qualitative result in all cases is independent on the actual sign of the parameter. For this reason, in the following, we will study DUNE's sensitivity to the modulus of these parameters, $|a_{\alpha\beta}|$, marginalizing over all possible phases, $\phi_{\alpha\beta}$.
The effect of the diagonal parameters\footnote{Note that one can always subtract a matrix proportional to the identity to the Hamiltonian, with no observable physical consequences. Here we have chosen to subtract the parameter $a_{\tau\tau}$ from the diagonal of the CPT--violating Hamiltonian in Eq.~(\ref{eq:cpt_part}), and therefore we redefine $\aee-\att$ and $\amm -\att$ as
$\aee$ and $\amm$, respectively.}, $\aee$  and   $\amm$, on  $\pme$ is  shown in Fig.~\ref{fig:prob_e_aii}. 
There, we observe that the impact of $\aee$ on the oscillation probability is much more noticeable than the effect of $\amm$.  
Note also that the sign of the former parameter determines the increase or decrease on $\pme$.

\begin{figure}[t!]
\centering
\includegraphics[scale=0.53]{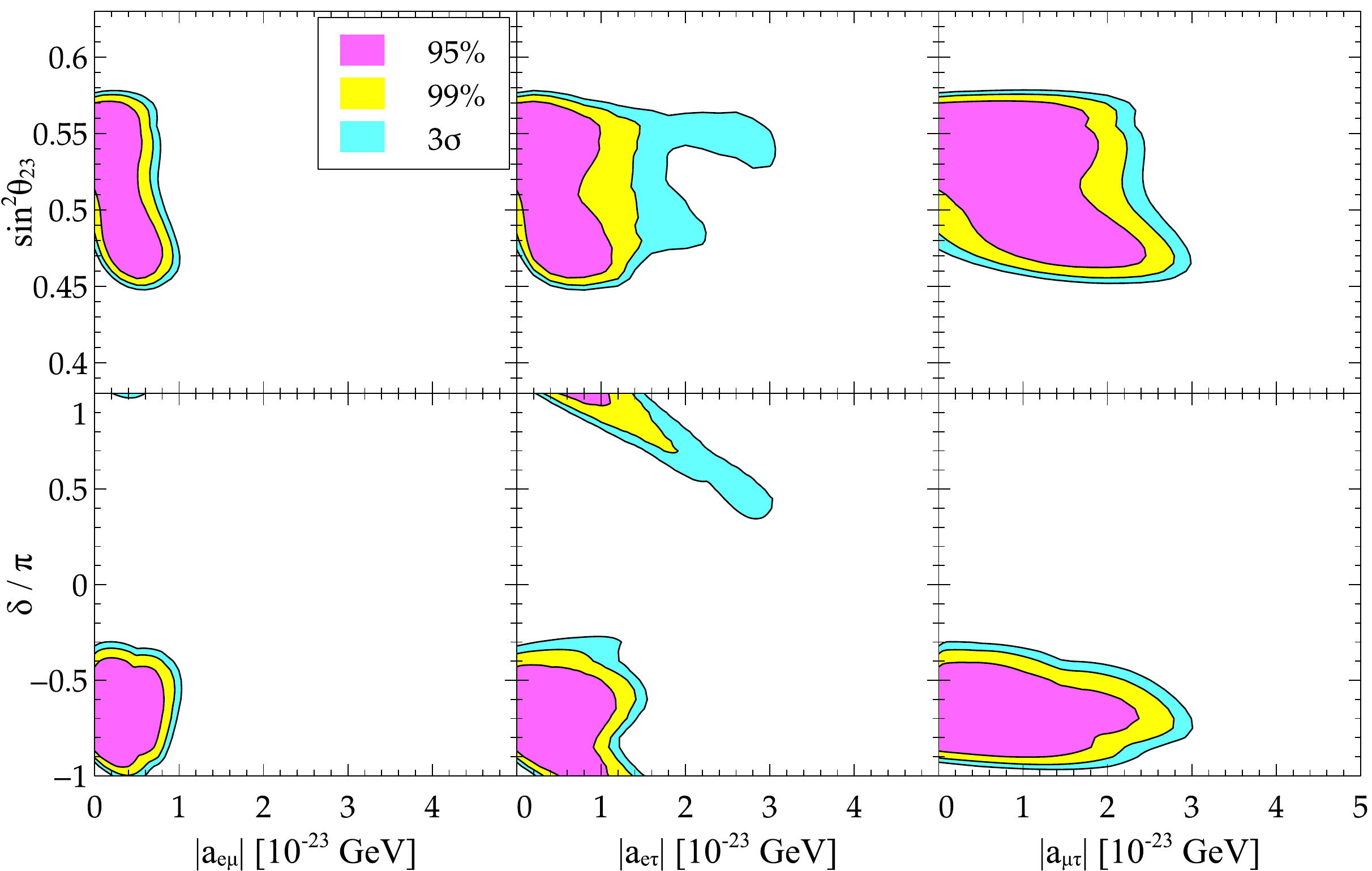}
\caption{\footnotesize{DUNE sensitivity to the simultaneous probe of the non--diagonal CPT--violating parameters $|\aem|$, $|\aet|$ and $|\amt|$  and the oscillation parameters $\theta_{23}$ and $\delta$ at $95\%$, $99\%$ and $3\sigma$ C.L.}}
\label{fig:aij_d_t23}
\end{figure}

\begin{figure}[t!]
\centering
\includegraphics[scale=0.53]{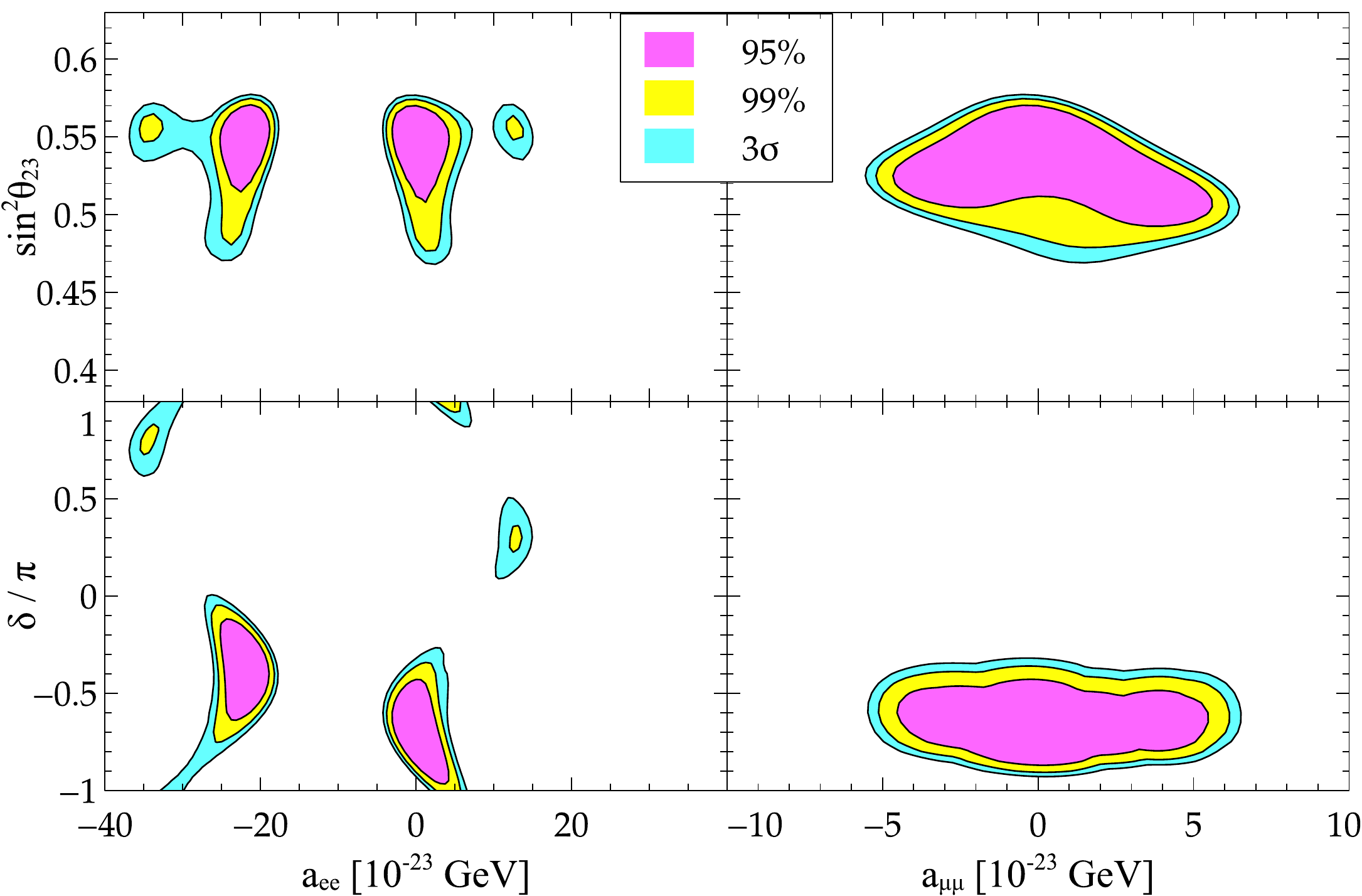}
\caption{\footnotesize{DUNE sensitivity to the diagonal CPT--violating parameters $\aee$ and $\amm$ and the oscillation parameters $\theta_{23}$ and $\delta$ at $95\%$, $99\%$ and $3\sigma$ C.L.}}
\label{fig:aii_d_t23}
\end{figure}

\begin{figure}[h!]
\centering
\includegraphics[scale=0.53]{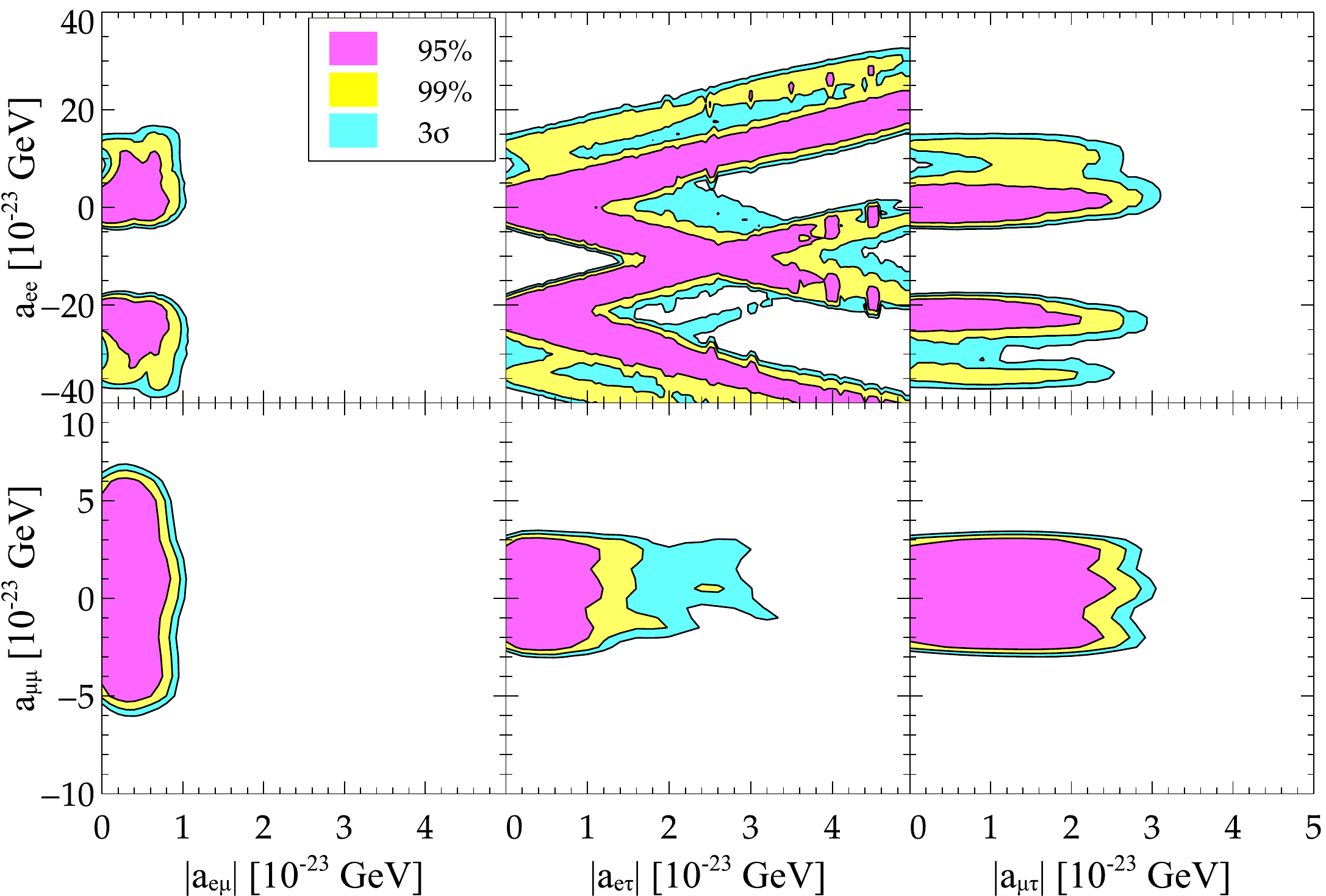}
\caption{\footnotesize{Correlations between the non--diagonal ($|\aem|, |\aet|, |\amt|$) and  diagonal ($\aee, \amm$) CPT--violating parameters at DUNE. The color code is the same as in Figs.~\ref{fig:aij_d_t23} and \ref{fig:aii_d_t23}.}}
\label{fig:aij_aii}
\end{figure}

\begin{figure}[h!]
\centering
\includegraphics[scale=0.45]{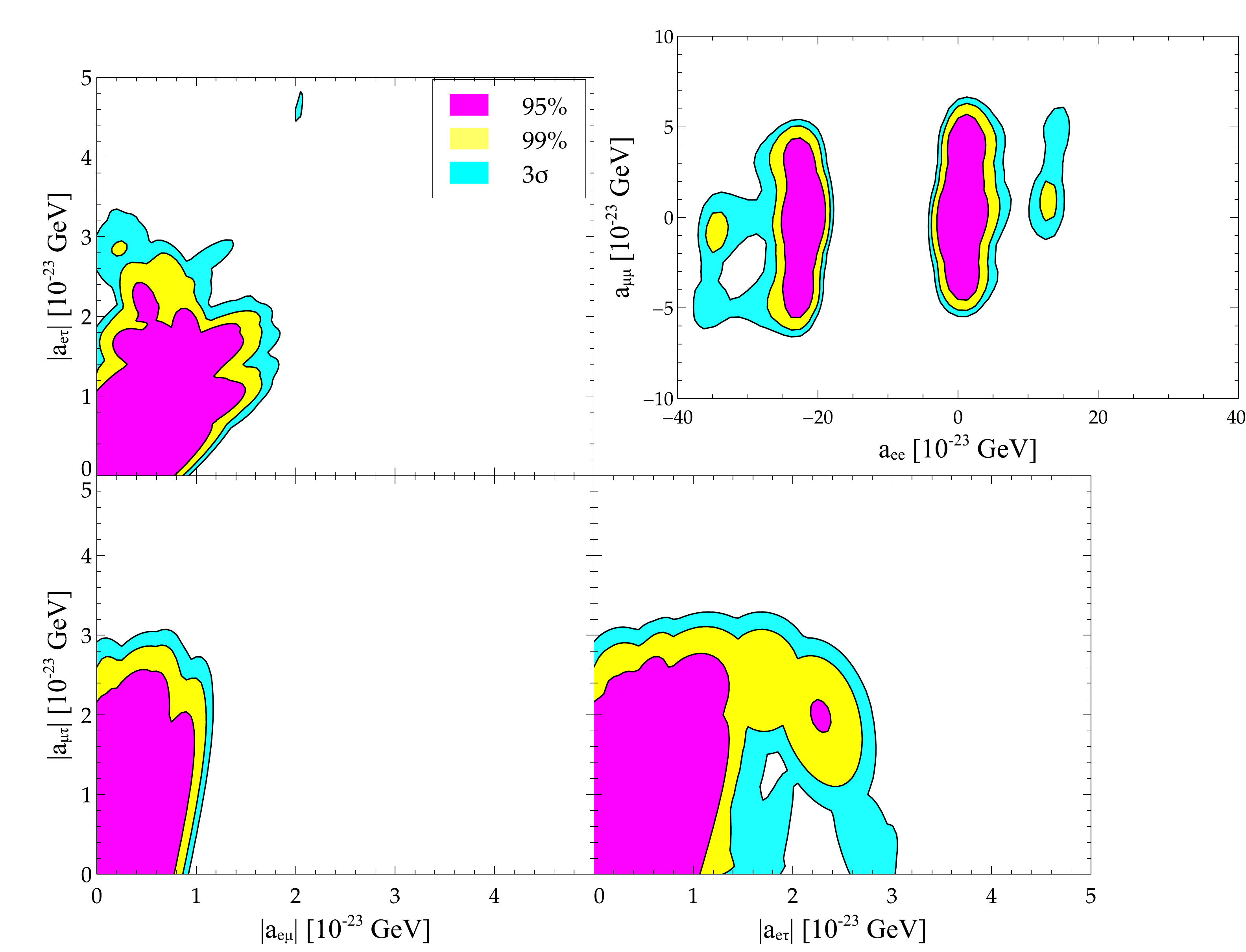}
\caption{\footnotesize{Expected correlations from our DUNE simulation among the non--diagonal CPT--violating  parameters, $|\aem|, |\aet|, |\amt|$ (left), and the  diagonal ones, $\aee$ and $\amm$ (right). The color code is the same as in previous figures.}}
\label{fig:cptv_corr}
\end{figure}

The main results of our sensitivity analysis of DUNE to CPT violation are  summarized in Figs.~\ref{fig:aij_d_t23} to \ref{fig:cptv_corr}. 
In these figures we plot the allowed regions at  $95\%$, $99\%$ and $3\sigma$ C.L. in the two--dimensional planes defined by considering two non--zero parameters among
 $\delta_{CP}$, $\theta_{23}$, $a_{ee}$, $a_{\mu\mu}$, $|a_{e\mu}|$, $|a_{e\tau}|$ and $|a_{\mu\tau}|$,  at a time.
 These two--dimensional regions indicate future DUNE's sensitivity  to constrain signals of Lorentz violation assuming as well DUNE's  future sensitivity to the  neutrino oscillation parameters.

 As we can see from Figs.~\ref{fig:aij_d_t23} and \ref{fig:aii_d_t23}, some degeneracies appear between the CPT--violating coefficients $a_{\alpha\beta}$ and the standard oscillation parameters $\sstc$ and  $\dcp$, mostly due to correlations with the atmospheric mixing angle that will not be solved by future DUNE data.
From the top row of Fig.~\ref{fig:aij_d_t23}, we see that the presence of  non--zero off--diagonal parameters $a_{\alpha\beta}$  can  give rise to degenerate regions corresponding to the opposite octant solution\footnote{As indicated in Table \ref{tab:oscpar}, we are assuming the true value of the atmospheric angle $\theta_{23}$ in the second octant.}.
Such a degeneracy is not present when allowing for non--zero diagonal coefficients $a_{\alpha\alpha}$ and, therefore, the most favoured region  in Fig.~\ref{fig:aii_d_t23} remains in the assumed true octant for the atmospheric angle.
 Concerning the simultaneous measurement of the CP phase  $\delta$  and the Lorentz--violating coefficients, from  Figs.~\ref{fig:aij_d_t23} and \ref{fig:aii_d_t23} we see that the determination of $\delta$ is more affected by the parameters that are more relevant for $P_{\mu e}$, $a_{e\tau}$ and $a_{ee}$, as seen in the probability plots in Figs.~\ref{fig:prob_e_aij} and \ref{fig:prob_e_aii}.
Then, although values of $\delta$ in the range ($\pi$/2, $\pi$) are only allowed with confidence levels above 99\%, it is clear that the proposed extremely good sensitivity to CP violation at DUNE can be compromised by the presence of Lorentz violation in the neutrino propagation.
Additionally, we also note in Fig.~\ref{fig:aii_d_t23} that new degenerate regions appear 
in presence of $a_{ee}$ (around $a_{ee} \sim -22 \times 10^{-23}$ GeV). This new degeneracy in $a_{ee}$ appears due to the marginalization over the two neutrino mass orderings.

On the other hand, the  correlation between the diagonal and the non--diagonal CPT--violating parameters is shown in  Fig.~\ref{fig:aij_aii}.
From the top central panel, one can notice a clear correlation  between $|\aet|$ and $\aee$, that can be explained from the impact of such parameters on the appearance electron neutrino probability $\pme$. Indeed, in Figs.~\ref{fig:prob_e_aij} and \ref{fig:prob_e_aii}, we see that a positive (or negative) value of $\aet$   and $\aee$ shifts the oscillation probability in opposite directions and, therefore, the two effects can compensate each other giving rise to degenerate solutions. Note that, in this figure, we are plotting the modulus of the complex off--diagonal parameters, $|a_{\alpha\beta}|$, while marginalizing over the  phases associated to them.  Therefore,  one of the two branches (originating approximately from $a_{ee} \sim 0$) in the plot under discussion corresponds to negative values of $a_{e\tau}$. 
In the same panel, two more branches stem from approximately $a_{ee} \sim -22 \times 10^{-23}$ GeV. This additional degeneracy in the presence of $a_{ee}$ can be observed in the other two panels of the top row of Fig.~\ref{fig:aij_aii}, too.
Besides this one, no other particular parameter degeneracies between the diagonal and non--diagonal Lorentz--violating coefficients can be observed in  Fig.~\ref{fig:aij_aii}.

Finally,  Fig.~\ref{fig:cptv_corr} shows the allowed regions in the two-dimensional planes delimited by the  three non--diagonal CPT--violating parameters (left) and the two diagonal ones (right panel).
 The most relevant result here is the correlation between $|\aet|$ and   $|\aem|$ that, in fact, is more apparent for the $95\%$ C.L. contour. This  originates from the similar, but opposite in magnitude, role of $|\aem|$ and $|\aet|$ in the main appearance channel $\nu_{\mu} \to \nu_{e}$ (see Fig.~\ref{fig:prob_e_aij}). 
The degeneracy corresponding to other than the true solution of $a_{ee}$ due to marginalization over the mass ordering is also apparent in the top-right panel of Fig.~\ref{fig:cptv_corr}.

%%%%%%%%%%%%%%%%%%%
\section{Summary and conclusions}
\label{sec:summary}
%%%%%%%%%%%%%%%%%%%

In this section we summarize the expected sensitivity to the CPT--violating parameters at the future DUNE experiment, comparing our results with the existing  bounds in the literature.

Several studies have been performed on this topic in the context of  the long--baseline experiment MINOS~\cite{Adamson:2012hp} and the reactor experiment Double Chooz~\cite{Abe:2012gw}. However, the most stringent bounds on the CPT--violating parameters  have been determined by the analysis of Super--Kamiokande atmospheric data in Ref.~\cite{Abe:2014wla}. Given the large distances travelled by atmospheric neutrinos, ranging up to $\sim$13000 km, the diameter of the Earth, the neutrino atmospheric flux offers an optimal scenario to test the violation of Lorentz invariance.
The limits obtained by the Super--Kamiokande collaboration are listed in the central column of Tab.~\ref{tab:bounds}. 
Note, however, that only direct limits on the off--diagonal parameters have been reported.

\begin{table}[t!]
\centering
\begin{tabular}{ccc}
\hline \\[-3mm]  
Parameter & \hphantom{~~}Existing Bounds\hphantom{~~} & This work  \\   \hline  \\[-3mm]
$|\aem|$ [GeV] & $2.5 \times 10^{-23}$~\cite{Abe:2014wla} & $7.0 \times 10^{-24}$ \\ \\[-3mm]  
$|\aet|$ [GeV]& $5.0 \times 10^{-23}$~\cite{Abe:2014wla} & $1.0 \times 10^{-23}$ \\ \\[-3mm]  
$|\amt|$ [GeV]& $8.3 \times 10^{-24}$~\cite{Abe:2014wla} & $1.7 \times 10^{-23}$ \\ \\[-3mm] 
\aee [GeV]& -- & $-2.5 \times 10^{-22} < \aee < -2.0 \times 10^{-22}$ and $ -2.5 \times 10^{-23} < \aee < 3.2 \times 10^{-23}$  \\ \\[-3mm]  
\amm [GeV]& -- & $ -3.7 \times 10^{-23} < \amm < 4.8 \times 10^{-23}$   \\   \hline
\end{tabular}
\caption{\footnotesize{Comparison between the 95\% C.L. existing bounds on CPT--violating parameters and the limits estimated in this work from the simulation of DUNE, as obtained from Fig.~\ref{fig:dchi2-aij}. 
}}
\label{tab:bounds}
\end{table}

\begin{figure}[t!]
\centering
\includegraphics[width=0.92\textwidth]{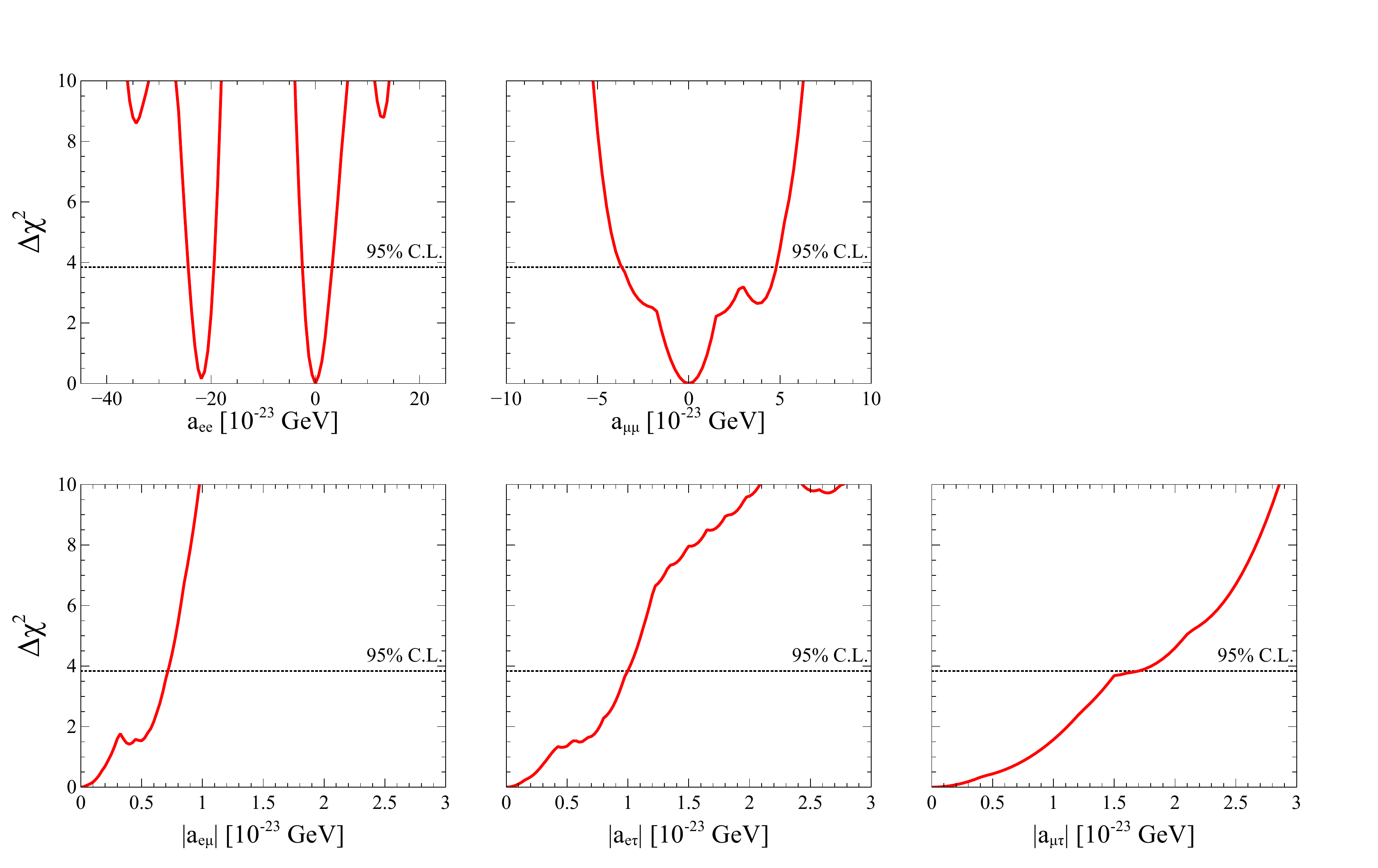}
\caption{\footnotesize{Expected DUNE sensitivity to the CPT--violating parameters $a_{\alpha\beta}$. The dashed line indicates the 95\% C.L. limit.}}
\label{fig:dchi2-aij}
\end{figure}

DUNE's  sensitivity to the Lorentz--violating parameters $a_{\alpha\beta}$ is presented in Fig.~\ref{fig:dchi2-aij}, where we plot the corresponding $\Delta\chi^2$ profiles after the marginalization over the standard oscillation parameters. These results are also presented as 95\% C.L. bounds on the CPT--violating coefficients in the right column of Tab.~\ref{tab:bounds}. 
We note that DUNE will be able to improve the current limits on the most  relevant CPT--violating parameters for the neutrino appearance channel, $\aem$ and $\aet$, by a factor 4--5.
Unfortunately,  DUNE's poor sensitivity to oscillations in the $\nu_\mu \to \nu_\tau$ channel will not allow improving the current limit on  $\amt$.
As commented before, no direct limits on the diagonal CPT--violating parameters $\aee$ and $\amm$ have been derived in the literature until now. Therefore, our estimated results in Table \ref{tab:bounds} are the first ones derived explicitly under the assumption of Lorentz violation.
We see that DUNE will be able to provide constraints on these parameters of the same order of magnitude as for the off--diagonal ones.
Here we can also remark the presence of an additional allowed region for $a_{ee}$  that, contrarily to all the other regions found, is not centered in zero. This solution appears 
due to an intrinsic degeneracy between the neutrino mass ordering and the LIV parameter $a_{ee}$, as discussed above. Such degeneracy is analogous to the one existing in the NSI scenario~\cite{Miranda:2004nb,Coloma:2016gei} and it affects simultaneously all the neutrino oscillation experiments. As for the NSI case, complementary measurements would be needed in order to resolve this degeneracy.\\

So far we have only considered the limits coming from dedicated searches of Lorentz violation at neutrino experiments. However, as proposed in Ref.~\cite{Diaz:2015dxa}, one can exploit the approximate equivalence  between the NSI and Lorentz--violating scenarios to translate the current bounds on NSI into limits on the $a_{\alpha\beta}$ parameters.
To do so one can consider 
\be\label{eq:nsi_cpt2}
a_{\alpha\beta} \equiv \sqrt{2}G_{F} N_{e}\epsilon^{m}_{\alpha\beta}\, ,
\ee
assuming an average matter density along the neutrino path. For an approximate value of the matter density $\rho$ $\sim$ 3.4 g/cm$^3$~\cite{Diaz:2015dxa},
and considering the existing NSI bounds from Super--Kamiokande atmospheric data~\cite{Mitsuka:2011ty}\footnote{The following bounds have been derived in the two--neutrino approximation.
A full three--flavour neutrino approach results in  looser limits~\cite{Friedland:2004ah,GonzalezGarcia:2011my}.}:
\begin{equation}
|\epsilon_{\mu\tau}^{dV}| < 0.011 \, ,\quad   |\epsilon_{\mu\mu}^{dV}-\epsilon_{\tau\tau}^{dV}| < 0.049 \, \quad \rm{(90\% \, C.L.)}\, ,
\end{equation}
one can derive the following approximate limits on the Lorentz--violating parameters\footnote{Note that there is a difference of factor of 3 with respect to the results obtained in Ref.~\cite{Diaz:2015dxa}, due to a more correct interpretation of the NSI bounds here. See Ref.~\cite{Farzan:2017xzy} for more details.}
\begin{equation}
|a_{\mu\tau}| < 4.3\times10^{-24} {\text{ GeV}} \, ,\quad   |a_{\mu\mu}-a_{\tau\tau}| < 1.9\times10^{-23}   {\text{ GeV}} \, \quad \rm{(90\% \, C.L.)}\, .
\end{equation}

Currently, the strongest constraints on NSI in the $\mu-\tau$ sector have been derived using high--energy atmospheric data from IceCube~\cite{Salvado:2016uqu}\footnote{{See Ref.~\cite{Esteban:2018ppq} for a recent  reanalysis of neutrino NSI from global oscillation data. This work adopts a more generic assumption about neutrino NSI with quarks and, as a result, a softer bound on the NSI  $\mu-\tau$ coupling is obtained.}}
\be
- 0.006 < \epsilon_{\mu\tau}^{dV} < 0.0054 \, \, \quad \rm{(90\% \, C.L.)}\, .
\ee
From these results one can derive new approximate limits on the $\mu-\tau$ CPT--violating parameter
\begin{equation}
-2.3\times10^{-24} {\text{ GeV}} < a_{\mu\tau} < 2.1\times10^{-24} {\text{ GeV}} \, ,
\end{equation}
improving the previous Super--Kamiokande bound by a factor of 2.

Likewise, considering the future sensitivity on neutrino NSI from the PINGU experiment~\cite{Choubey:2014iia}, one can translate it into projected limits on the CPT--violating parameters
\begin{eqnarray}
-5.6\times10^{-25}  {\text{ GeV}} < & a_{\mu\tau} & <  6.1\times10^{-24}   {\text{ GeV}}\, , \nonumber \\
 -3.9\times10^{-24}  {\text{ GeV}}<  &(a_{\mu\mu}-a_{\tau\tau}) & <  2.2\times10^{-24} {\text{ GeV}} \, \quad \rm{(90\% \, C.L.)}\, .
\end{eqnarray}

Note, however, that this translation between NSI and Lorentz--violating limits lies on the approximation of assuming an average density along the neutrino path, not very accurate in the case of the atmospheric neutrino flux.

%%%%%%%%%%%%%%%%%%%
\section*{Parting thoughts}
%%%%%%%%%%%%%%%%%%%

%Clearly, as we have shown, 
Massive neutrinos, who were the last addition to the Standard Model, and probably the
first piece of evidence of the physics that hides behind it, have become an ideal tool to test the foundations of the
way we understand and describe Nature, local relativistic quantum field theory.

%Clearly, as we have shown, massive neutrinos have become an ideal tool to test the foundations of the way we understand and describe Nature  using local relativistic quantum field theory. 

Quantum gravity effects which are suppressed by the Planck scale, effects of the high scale associated to neutrino masses
if neutrinos are Majorana particles, non--local effects due to string scales, and effects we can nowadays only dream about will be measured or bounded in the near future. If it results in a modification to the dispersion relation, it leaves a trace in the oscillation probability that can be searched for.
Long--baseline neutrino experiments involving very large distances are the magnifying glass that can enhance a miserably small effect $a \sim 10^{-24}$ GeV to a difference in the transition probability that can be even seen by eye!
In this respect, neutrinos are the most accurate tool we have to test, not the Standard Model itself, but the building blocks we used to construct it. Neutrinos will not challenge the building, they can challenge the construction system. A fact that should not be underestimated.

%%%%%%%%%%%%%%%%%%%
\section*{Acknowledgements}
%%%%%%%%%%%%%%%%%%%

GB acknowledges support from the MEC and FEDER (EC) Grants SEV-2014-0398, FIS2015-72245-EXP, and FPA-2017-84543P  and the Generalitat Valenciana under grant PROMETEOII/2017/033. GB acknowledges partial support from the European Union FP7 ITN INVISIBLES MSCA PITN-GA-2011-289442 and InvisiblesPlus (RISE) H2020-MSCA-RISE-2015-690575. 
CAT, MM and MT are  supported by the Spanish grants FPA2017-85216-P and
SEV-2014-0398 (MINECO) and PROMETEO/2018/165 (Generalitat Valenciana). 
CAT is supported by the FPI fellowship BES-2015-073593 (MINECO).
MT acknowledges financial support from MINECO through the Ram\'{o}n y Cajal contract RYC-2013-12438 as well as from the L'Or\'eal-UNESCO \textit{For Women in Science} initiative.

\bibliographystyle{apsrev}

%\bibliography{bibliography.bib}

\end{document}